\documentclass[pre,twocolumn,showpacs,amsmath,amssymb,10pt,a4paper,nofootinbib]{revtex4}

\usepackage{geometry}
\usepackage{graphicx}
\usepackage{dcolumn}
\usepackage{bm}
\usepackage{subfigure}
\usepackage{amsmath}
\usepackage{amsfonts}
\usepackage{multirow}
\newcommand{\be}{\begin{equation}}
\newcommand{\ee}{\end{equation}}
\newcommand{\bd}{\begin{displaymath}}
\newcommand{\ed}{\end{displaymath}}
\newcommand{\BE}{\begin{eqnarray}}
\newcommand{\EE}{\end{eqnarray}}

\newcommand{\id}{{\openone}}

\newcommand{\bomega}{{\mbox{\boldmath $\omega$}}}

\newcommand{\bsigma}{{\mbox{\boldmath $\sigma$}}}
\newcommand{\blambda}{{\mbox{\boldmath $\lambda$}}}
\newcommand{\bmu}{{\mbox{\boldmath $\mu$}}}
\newcommand{\bnu}{{\mbox{\boldmath $\nu$}}}

\newcommand{\bra}[1]{\ensuremath{\langle#1|}}

\newcommand{\ket}[1]{\ensuremath{|#1\rangle}}

\newcommand{\ketbra}[1]{\ensuremath{| #1 \rangle \langle #1 |}}

\newcommand{\NN}{\ensuremath{\mathcal{N}}}
\newcommand{\LL}{\ensuremath{\mathcal{L}}}
\newcommand{\TT}{\ensuremath{\mathcal{T}}}

\newcommand{\QQ}{\ensuremath{\mathcal{Q}}}

\newcommand{\EEE}{\ensuremath{\mathcal{E}}}

\begin{document}

\title{Complexity measures, emergence, and multiparticle correlations}

\author{Tobias Galla}
\email{tobias.galla@manchester.ac.uk}
\affiliation{Theoretical Physics, School of Physics and Astronomy, The
University of Manchester, Manchester M13 9PL, United Kingdom}

\author{Otfried G\"uhne}
\email{otfried.guehne@uni-siegen.de}
\affiliation{Naturwissenschaftlich-Technische Fakult\"at, Universit\"at Siegen,
Walter-Flex-Stra{\ss}e 3, 57068 Siegen, Germany}
\affiliation{Institut f\"ur Quantenoptik und Quanteninformation,
\"Osterreichische Akademie der Wissenschaften, Technikerstra{\ss}e 21, 6020
Innsbruck, Austria}

\date{\today}

\begin{abstract}
We study correlation measures for complex systems. First, we investigate some
recently proposed measures based on information geometry. We show that these 
measures can increase under local transformations as well as under discarding 
particles, thereby questioning their interpretation as a quantifier for complexity 
or correlations. We then propose a 
refined definition of these measures, investigate its properties and discuss its
numerical evaluation. As an example, we study coupled logistic maps and study the behavior
of the different measures for that case. Finally, we investigate other local effects
during the coarse graining of the complex system.
\end{abstract}

 
\maketitle
\section{Introduction}
The science of complex systems is an active area of research in physics 
and in adjacent disciplines. Many systems in physics, biology, the social 
sciences and in economics fall under its umbrella, and research into these 
topics is becoming increasingly interdisciplinary. The quantification 
of complexity, however, is not striaghtforward and there are many approaches 
to quantify complexity in more mathematical terms, see for example
Refs.~\cite{crutchfield94,crutchfield89,baryam,chalmers,mackay05,mackay08, diakonova10,diakonova11} or
Ref.~\cite{lindgren88}, where a number of complexity measures are listed 
and discussed.

A possible way to quantify the complexity of a correlated multiparticle 
system is to take its distribution in state space and consider its distance 
to the distribution of an uncorrelated system, measured for example by the 
multi-information or excess entropy \cite{kahle09,ay}. As recently demonstrated, 
this concept can be generalised to higher-order multi-particle correlations 
using methods of information geometry \cite{amari01, kahle09, olbrich09}. For 
a given $N$-particle system one can consider the space ${\cal E}_k$ of 
$N$-partite distributions which are thermal states of $k$-particle Hamiltonians, 
and ask what the distance of a given $N$-particle distribution is from this space. 
These ideas were discussed in
detail in Ref.~\cite{kahle09}, and an appropriate mathematical definition in
terms of exponential families was devised. The authors of Ref.~\cite{kahle09}
have also made available an numerical algorithm with which to compute the
complexity measures they introduce \cite{cipi}.

While the above mentioned approaches deal with {\it classical} complex 
systems, correlations play also a vital role in {\it quantum} multi-particle
sysems. Consequently, many studies have been devoted to the characterization 
of quantum correlations \cite{linden02,reviews,zhou06,zhou08,zhou09}. Concepts 
from (classical) information geometry were extended to the quantum realm by Zhou
\cite{zhou08,zhou09}, in particular a quantum analog of the above exponential
families ${\cal E}_k$ was introduced. Based on these concepts the author of
Ref.~\cite{zhou09} arrives at the somewhat counterintuitive observation that
local operations on a single particle can increase the overall correlation in a
quantum state. More precisely, Zhou gives an explicit example of a thermal
quantum state of a two-particle Hamiltonian on three particles, where a local
operation on one of the three qubits 
converts the state into one which can no longer be written as a thermal state of
a two-particle Hamiltonian, but which instead has three-particle
correlations\footnote{In the following, statements such as `a state has
$k$-particle interactions' will always be taken to mean that the state cannot be
written as the thermal (Gibbs) state of a Hamiltonian with $k-1$-particle
interaction alone, but that instead any Hamiltonian generating the state
necessarily has at least one $k$-particle term. The absence of $k$-particle
correlations in this sense is not equivalent to a factorization requirement on
$k$-particle {\em correlation functions} often considered in statistical
physics.}.

One of the main motivations for the present work is the question whether or not
similar effects can be seen in classical complex systems. To this end we
characterize the class of local transformations, acting on single particles, and
investigate how multi-particle correlation, as defined in Ref.~\cite{kahle09} is
affected by those transformations. We show that (i) local transformations can
turn a distribution generated by a two-particle Hamiltonian into one with
genuine three particle correlation, and as a corollary (ii) that integrating out
individual particles can increase multi-particle correlation as well. 

In order to provide a remedy for these undesired properties we propose the
concept of a `local orbit' of an exponential family. The local orbit of a set
$S$ of distributions is here the set of all distributions that can be generated
from elements of $S$ by applying local transformations. We investigate the
possibility to define multi-particle complexity not in terms of distances from
exponential families themselves, but from their local orbits, and we propose a
numerical scheme with which these complexity measures can be approximated.
Finally we discuss a second type of local intervention, and study how different
coarse-graining procedures can affect the multi-particle correlation properties
of a system of coupled chaotic maps.

The remainder of this paper is structured as follows: In Sec.~\ref{sec:ig} we
provide the basic concepts of information geometry and briefly describe the
complexity measures introduced in Ref.~\cite{kahle09}. The focus of
Sec.~\ref{sec:lt} is the concept of local transformations. We first introduce
their mathematical definition, and then show how local transformations can
increase the correlation content of multi-particle systems. In Sec.~\ref{sec:im}
we propose an alternative set of complexity measures, invariant under local
transformations. We also describe a numerical method with which to approximate
these measures. In 
Sec.~\ref{sec:log} we apply these concepts to a system of coupled chaotic maps,
before we conclude in Sec.~\ref{sec:con}. Details of the numerical procedures
are outlined in the Appendix.

\section{Complexity measures from information geometry}\label{sec:ig}
In this section, we briefly review the complexity measures used in
Ref.~\cite{kahle09}. The underlying idea is the following: Consider a dynamical
system composed of $N$ particles each of which can be observed to be in one of
two different states, referred to as $0$ and $1$ in the following. At any given
time the state $\bsigma=(\sigma_1,\dots,\sigma_N)$ of the system is therefore
found to be an element of $\{0,1\}^N$. The space of all $2^N$ possible
configurations $\bsigma=(\sigma_1,\dots,\sigma_N)$ will be denoted as
$\Omega:=\{(\sigma_1,\dots,\sigma_N), \sigma_i\in\{0,1\},
i=1,\dots,N\}=\{0,1\}^N$. The stationary distribution of the system is then a
probability distribution over $\Omega$. 

Given such a distribution,  $P(\cdot)$, one can ask whether or not it is the
thermal state of $k$-particle Hamiltonian, $0\leq k\leq N$, i.e., whether one
can write $P(\bsigma)=Z^{-1} \exp[H^{(k)}(\bsigma)]$, where $H^{(k)}(\bsigma)$
is a Hamiltonian containing only $j$-particle terms with $j=1,\dots,k$, and
where $Z$ is a constant ensuring normalisation.  Trivially, any $N$-particle
distribution can be written as the Gibbs state of an $N$-particle Hamiltonian,
but for $k<N$ this may not be the case.  One can then use the distance $D_k$
(quantified by the  relative entropy) from the set of all distributions
generated by $k$-particle interactions as a measure of complexity. In order to
make this idea mathematically precise, we first give the definition of
exponential families, their closure, and the relative entropy. We here follow
the work and notation of Ref.~\cite{kahle09}. We then discuss some properties of
the resulting distance measures in Sec.~\ref{sec:dm}.

\subsection{Exponential families}

We will write the set of all particles as $V=\{1,\dots,N\}$ in the following. 
Given a subset $A\subset V$, we write $H_A(\cdot)$ for Hamiltonians (or, more
generally, arbitrary functions) which only depend on the $\{\sigma_i\}$ with
$i\in A$, i.e. $H_A=H_A(\{\sigma_i\}_{i\in A}).$ A $k$-particle Hamiltonian
($k\leq N$) is then a Hamiltonian containing only terms which each depend on (at
most) $k$ spins, so it is of the form \be H(\bsigma)=\sum_{A\subset V: |A|=k}
H_A(\{\sigma_i\}_{i\in A}).  \ee We note that any $k$-particle Hamiltonian is
also an $\ell$-particle Hamiltonian for all $\ell\geq k$. The vector space of
all $k$-particle Hamiltonians will be denoted by ${\cal Q}_k$ in the following,
and accordingly we have ${\cal Q}_k\subset {\cal Q}_\ell$ for all $\ell > k$.

The set of thermal states (or Gibbs measures) generated by $k$-particle
Hamiltonians then constitutes a so-called exponential family. We write
\be 
{\cal E}_k:=\left\{P | P(\bsigma)=\frac{e^{H(\bsigma)}}{\sum_{\bsigma'}
e^{H(\bsigma')}}, H\in {\cal Q}_k\right\}\subset{\cal P}(\Omega) , 
\ee 
where ${\cal P}(\Omega)$ denotes the set of all probability distributions over
$\Omega$. We then have 
\be {\cal E}_1\subsetneq {\cal E}_2\subsetneq \dots\subsetneq {\cal E}_N={\cal
P}(\Omega).  
\ee 
We will first discuss some of the properties of the families ${\cal E}_k.$
First, it is important to note that the probability distribution in ${\cal E}_1$
are simply product distributions factorizing over single particles, i.e. they
are distributions which can be written as $P(\bsigma)=P^{(1)}(\sigma_1)
P^{(2)}(\sigma_2) \cdot ...\cdot P^{(N)}(\sigma_N)$ with single-particle
distributions $P^{(i)}(\cdot)$.  The distributions in the more complex
exponential family ${\cal E}_2$ on the contrary can be written as products of
the form 
\be 
P(\bsigma) =\NN \prod_{{i,j \in V}, \; {i\neq j}} P^{(ij)}(\sigma_i,\sigma_j),
\ee 
where the $\{P^{(ij)}\}$ are two-particle distributions and where the constant
$\NN$ provides appropriate normalization. Factorizations of this type can be
extended straightforwardly to describe the elements of ${\cal E}_k$ as products
of $k$-particle distributions. There are other equivalent characterizations of
${\cal E}_k$, we will discuss and use some of them later in Section
\ref{localincrease}.

Given that Hamiltonians assign finite energies to all configurations $\bsigma$,
i.e. $H(\bsigma)\in\mathbb{R}$ for all $H\in{\cal Q}_k$, $k=0,\dots,N$ and all
$\bsigma\in\Omega$ the probability distributions ${\cal E}_k$  by construction
carry non-vanishing probability for all $\bsigma \in \Omega.$, i.e.
$P(\bsigma)>0$ for all $\bsigma$ and all $P\in\EEE_k$. It is therefore useful to
consider the closure $\overline{\cal {E}}_k$ to incorporate probability
distributions without full support.  For example, the three-particle probability
distribution $P$ with $P(000)=P(111)=1/2$ and $P(\cdot)=0$ elsewhere is not in
${\cal E}_2$, but in $\overline{\cal {E}}_2,$ since it can be approximated by
the low-temperature limit of an Ising-type two-particle interaction.  The
generalization of this distribution to more than three particles can still be
generated with two-particle interactions only. On the other hand, the
$N$-particle distribution $P$ with 
\be
\label{dickedistri}
P(\bsigma)=\left\{\begin{array}{ll} 1/2^{N-1} ~~ &~~ \mbox{if } \sum_i \sigma_i
\equiv 0 ~\mbox {mod} ~2, \\ ~& ~\\
0 ~~& ~~\mbox{elsewhere}\end{array}\right.
\ee
cannot be generated via $N-1$-particle interactions, but instead requires a
Hamiltonian with $N$-particle interaction. This distribution assigns probability
$2^{-(N-1)}$ to any bitstring $\bsigma$ with an {\em even} number of bits equal
to one, and zero to all other bitstrings. If only $N-1$ entries in $\bsigma$ are
known, it is not possible to tell whether or not this condition is met, hence
one cannot decide whether $P(\bsigma)=2^{-(N-1)}$ or whether $P(\bsigma)=0$,
this requires always knowledge of the state of all particles [see also point
(iii) below].

\subsection{Distance measures}\label{sec:dm}

Given a probability distribution $P(\cdot)$ over $\Omega$ and an interaction
order $k=0,\dots,N-1$, we then ask how closely $P(\cdot)$ can be approximated by
distributions generated by $k$-particle Hamiltonians. This is captured by the
following `distance' 
\be \label{eq:ddef}
D_k(P):=\inf_{Q\in\overline{\cal E}_k} D(P||Q), 
\ee 
where $D(P||Q)$ is the Kullback-Leibler distance \cite{coverthomas}
\be D(P||Q)=\sum_{\bsigma\in \Omega}
P(\bsigma)\log_2\frac{P(\bsigma)}{Q(\bsigma)}.  
\ee 
This quantity has the following properties: 
\begin{enumerate}

\item
Since $D(P||Q)=0$ if and only if $P=Q$, the quantity $D_k(P)$ 
is non-vanishing for all distributions $P\notin\overline{\cal E}_k.$ 

\item  The minimizing distribution ${Q^*\in\overline{\cal E}_k}$ on the RHS 
of Eq. (\ref{eq:ddef}) is given by the maximum likelihood approximation of $P$
by distributions in 
$\overline{\cal E}_k$, see Refs.~\cite{kahle09, amari01} for further details.  

\item
A second characterization of the minimizer $Q^*$ is the following: 
$Q^*$ is the distribution of maximal entropy in the set of distributions 
with the same $k$-particle marginals as $P$ \cite{amari01}.
This implies that the distribution from Eq.~(\ref{dickedistri}) has $D_k(P)=1$
for
all $k < N$, since the flat distribution has the same $k-1$-particle marginals 
as $P$ and clearly maximizes the entropy over all distributions.

\item For the case $k=1$ the minimizer $Q^*$ can directly be found as
\be
~~~~~~Q^*(\bsigma) = P^{(1)}(\sigma_1) P^{(2)}(\sigma_2) \cdot ...\cdot
P^{(N)}(\sigma_N),
\ee
where the $P^{(i)}(\cdot)$ are the single-particle marginals of $P.$ The 
quantity $D_1$ is also known as multi-information \cite{amari01,
multiinformation}, 
the above quantities $D_k$ can be considered as a generalization of this
concept.

\item For $k\geq 2$ an analytical calculation of $D_k(P)$ for a given
distribution $P\in{\cal P}(\Omega)$ is usually not straightforward.  There are,
however, powerful numerical tools for its computation \cite{cipi}, for
completeness we outline a possible algorithm in the Appendix.

\item Following Ref.~\cite{kahle09} one can also consider
\be
~~~~~I^{(k)}(P)=D_{k-1}(P)-D_k(P), k=1,\dots,N
\ee
as complexity measures. The quantity $D_k(P)$ here represents the {\em
improvement} in approximating $P\in{\cal P}(\Omega)$ when $k$-particle terms in
the generating Hamiltonian are allowed over the case in which only
$k-1$-particle interaction is admitted. In this paper, however, we will mainly
work with $D_k(P)$.

\item Finally, possible generalizations to the quantum setting have been
discussed in 
Refs.~\cite{linden02, zhou08, zhou09}.

\end{enumerate}

We also note that the choice of the Kullback-Leibler divergence
[Eq.~(\ref{eq:ddef})] is of course not the only choice of an underlying distance
between probability distributions. Other distance measures are conceivable, see
for example Refs.~\cite{mackay05,mackay08,diakonova11}. In our work we will
however restrict   the discussion to the relative entropy, as this provides
several useful properties of $D(\cdot||\cdot)$ which we will use later on.

\section{Local transformations}\label{sec:lt}

\begin{figure}[t]
\centering
\includegraphics[width=\columnwidth]{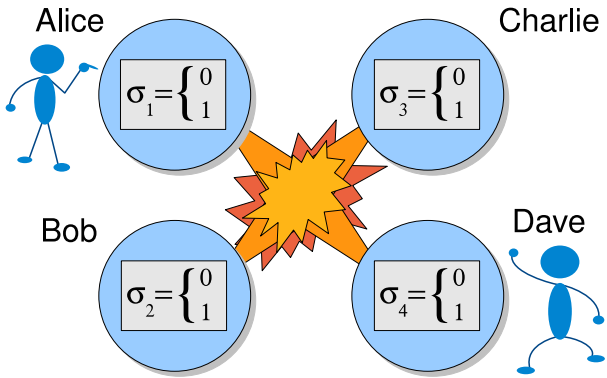}
\caption{(color online). Illustration of the effect of local transformations: 
A complex interacting dynamics generates a probability distribution over the
state space 
of four binary particles. Each observer (Alice, Bob, Charlie and Dave) has
access to the state of one particle, and reports these to a central authority
(not shown). One type of local transformation arises, if one party, say Alice,
changes the information she transmits stochastically, but without communication
with the other observers (see text for further details). A second possible type
of local transformation is given by disregarding one party, say Dave, and
considering only the marginal distribution on the remaining three parties (this
corresponds to a case in which Dave randomly reports to have seen state $0$ or
$1$ regardless of what he actually observed). It is natural to require from a
measure of complexity that it should not increase under either type of local
transformations.}
\label{fig:localfig}
\end{figure}

\subsection{Definition and interpretation}
We will next discuss the behavior of the measure $D_k$ under local 
transformations. In considering local transformations we have the following
scenario in 
mind (see also Fig.~\ref{fig:localfig}): a complex system composed of $N$
binary 
particles generates successive states
$\bsigma(t)=[\sigma_1(t),\sigma_2(t),\dots, \sigma_N(t)]$, 
where $t=1,2,...$ denotes time. The resulting stationary
distribution\footnote{We here only consider interacting systems which do have a
stationary distribution.} over $\Omega=\{0,1\}^N$ 
is then given by the cumulative histogram of all observed $N$-particles states
in the 
long-time limit, i.e. $P(\bsigma)=\lim_{t\to\infty} t^{-1}\sum_{t'=1}^t
\delta_{\bsigma\bsigma(t')}$, where $\delta$ is the standard Kronecker symbol,
i.e. $\delta_{\bsigma\bsigma'}=1$ if $\bsigma=\bsigma'$ and
$\delta_{\bsigma\bsigma'}=0$ otherwise. Computing this distribution however
requires a centralized observer, 
who has access to the sequence $\bsigma(t)$. 

To introduce the concept of local transformations we now consider a different
scenario 
and assume that there are $N$ distant observers who each have access to sequence
of 
states of only one of the $N$ particles, and then report these to a centralized 
agent who aggregates the information. By local transformations we mean
manipulations 
a single observer can execute on the state of the particle he (or she) has
observed 
before passing on the information to the centralized site. Assume for example
that 
observer $i\in\{1,\dots,N\}$  upon observation of state $0$ reports to have seen
state 
$0$ with probability $1-a_i$ and that he (or she) reports state $1$ with
probability 
$a_i$ in such circumstances. Similarly if presented with observation $1$ he 
or she may report state $1$ to the central site with probability $1-b_i$, but
pass 
on wrong information with rate $b_i$ ($a_i,b_i\in[0,1]$). The central site then
accumulates 
the information and obtains an $N$-particle  distribution $\widetilde
P(\cdot)$, 
computed from the reported data of the $N$ observers. 

Unless $P$ itself is pathological 
one will generally have $P=\widetilde P$ only if $a_i=b_i=0$ for all $i$.
Behaviour 
on the part of the observers in which this is not the case can be thought 
of as noise (i.e. inaccurate observation) or as deliberate falsification of 
the observed input. The $\{a_i,b_i\}$ are error rates, quantifying how
frequently inaccurate information is transmitted to the central agent. The
crucial point is that such manipulations occur locally, 
i.e. different observes may not collude with each other and communicate before 
reporting to the central site. All operations therefore act on single particles
only. 
It appears reasonable to ask that such transformations should not increase the 
complexity of $P(\cdot)$: if under a given measure of complexity $k$-particle 
correlations are absent in $P(\cdot)$ then applying local transformations 
should not introduce $k$-particle correlation in $\widetilde P$ (the effective
distribution perceived by the aggregating central agent). In a similar vein
disregarding one or several particles should not introduce additional
correlation: if one considers the marginal probability distribution of the
states of a subset of $M<N$ 
particles, one may require that these $M$-particle marginals should be not more
complex than the joint probability distribution of all $N$ particles. Similar
considerations are well known from quantum information theory, where a basic
requirement for entanglement measures is that they should not increase under
local operations and classical communication. 

In order to investigate whether the complexity measures defined above do indeed
respect these constraints we write $A^{(i)}_{\mu_i,\nu_i}$ for the probability
with which observer $i$ reports state $\mu_i\in\{0,1\}$ to the central site when
he/she has in fact seen state $\nu_i\in\{0,1\}$. We always have $\sum_{\mu_i}
A^{(i)}_{\mu_i,\nu_i}=1$ for all $i$ and $\nu_i$. We will use the notation 
$A^{(i)}={1-a_i\;\; b_i\choose a_i \; 1-b_i}$ in the following. A specific local
transformation is then characterized by the parameters
$\bomega=(a_1,b_1,\dots,a_N,b_N)\in[0,1]^{2N}$, we will often use $T_\bomega$ as
a short-hand for the transformation defined by $\bomega$.

For any local transformation $T^{\rm loc}=T_\bomega$ the resulting distribution
$\widetilde P(\cdot)$ put together by the central site can then be written as
\be
\widetilde P (\bmu) = \sum_{\bnu} T^{\rm loc}_{\bmu \bnu} P(\bnu),
\ee
where
\be \label{eq:tprod1}
T^{\rm loc}= \bigotimes_{i=1}^{N} A^{(i)}=\bigotimes_{i=1}^N
\left(\begin{array}{cc} 1-a_i & b_i\\ a_i & 1-b_i\end{array}\right).
\ee
 Equivalently the probability of the central site constructing the $N$-particle
configuration $\bmu$ from the reports he or she receives from the $N$ observers
when the true configuration of the $N$-particle system was $\bnu$ is given by
 \be\label{eq:tprod2}
 T^{\rm loc}_{\bmu \bnu} = \prod_{i=1}^{N} A^{(i)}_{\mu_i \nu_i}.
 \ee
The superscript `loc' here indicates that we are interested in local
transformations, i.e. those for which the matrix $T$ has the product structure
indicated in Eqs. (\ref{eq:tprod1}, \ref{eq:tprod2}). More generally we can
consider linear transformations $Q(\bmu)=\sum_\bnu T_{\bmu\bnu}P(\bnu)$. These
are well defined so long as the entries of the matrix $T_{\bmu \bnu}$ are
nonnegative and the entries in the columns add up 
to one, $\sum_\bmu T_{\bmu \bnu} =1$, i.e. so long as the matrix $T$ is
stochastic\footnote{If one requires in addition that the row sums are
normalized, $\sum_\nu T_{\bmu \bnu} =1$ 
the matrix is called doubly stochastic and the entropy increases during the
process.  Doubly stochastic matrices naturally occur, if one requires that the
maximally mixed distribution is invariant under the transformation.}.

In order to probe the measures of correlations introduced earlier we next
investigate how the distances $D_k$ behave under local operations.  As it turns
out, distances from exponential families can indeed increase under local
transformations.

\subsection{Local operations can increase $D_k$}
\label{localincrease}
To give an example, first note that under arbitrary transformations (not
necessarily local) the Kullback Leibler distance can only decrease, i.e.,
$D(P||Q) \geq D[T(P)||T(Q)]$ \cite{coverthomas}, where we have written $T(P)$
for the image of $P(\cdot)$ under a transformation mediated by the matrix $T$
[and similar for $T(Q)$]. As a consequence, the following two statements are
equivalent:
\begin{enumerate}
\item[(1)] $D_k$ decreases under local transformations, i.e. $D_k[T^{\rm
loc}(P)]\leq D_k(P)$ for all $P$ and all local transformations $T^{\rm loc}$, 
\item[(2)] the manifold $\EEE_k$ is invariant under local transformations, i.e.
$T^{\rm loc}(Q)\in\EEE_k$ for all $Q\in \EEE_k$ and all $T^{\rm loc}$.
\end{enumerate}
As we will now show neither of these two statements is true however.

In order to construct a counterexample it is useful to detail further the
characteristics of probability distributions in $\EEE_k$. For simplicity, we
focus on  the case of three particles and consider $\EEE_2$, generalization to
the other cases is straightforward. It turns out to be helpful to formulate the
problem in the language of quantum mechanics. In analogy to the space consisting
of the $2^3$ configurations $\{\bsigma\}=\{(\sigma_1, ... \sigma_3)\}= \{
(0,0,0), (0,0,1), ..., (1,1,1)\}$ we consider analogous quantum states in the
orthonormal computational basis, $\{\ket{\bsigma}\}=\{\ket{000}, \ket{001},...,
\ket{111}\}.$  For a given probability distribution $P(\bsigma)$ we then
consider the Hamiltonian defined by 
\be\label{eq:diagh}
H=\sum_\bsigma \lambda_\bsigma \ketbra{\bsigma},
\ee
where the $\{\lambda_\bsigma\}$ are given by $\lambda_\bsigma=\ln P(\bsigma)$.
For the moment we restrict the discussion to probability distributions $P$ with
full support, i.e. we assume $P(\bsigma)>0$ for all $\bsigma$. The operator $H$
is then well defined and diagonal in the computational basis,
$H\ket{\bsigma}=\lambda_\bsigma\ket{\bsigma}$, and we have
\be\label{eq:p}
 P(\bsigma)=\bra{\bsigma}\exp(H)\ket{\bsigma}.
 \ee
Using the fact that tensor products of Pauli matrices form a basis of the  space
hermitian operators,
any Hamiltonian of the type defined in Eq.~(\ref{eq:diagh}) can uniquely be
written in the form
\BE
H&=& \kappa\id+\sum_i \alpha_i \sigma_z^{(i)} + \sum_{i,j} \beta_{ij}
\sigma_z^{(i)}\otimes\sigma_z^{(j)} \nonumber \\
&&+ \gamma \sigma_z^{(1)}\otimes \sigma_z^{(2)}\otimes
\sigma_z^{(3)},\label{eq:alpha}
\EE
where $\sigma_z^{(i)}$ denotes the Pauli matrix $\sigma_z$ acting on qubit $i$. 
We will in the following generally not treat $\kappa$ in Eq.~(\ref{eq:alpha}) as
free parameter, but choose
it such that the distribution $P(\cdot)$ constructed from $H$ via
Eq.~(\ref{eq:p}) is correctly normalised. 

A Hamiltonian of the type described in Eq. (\ref{eq:alpha}) is a two-particle
Hamiltonian if and only if the $3$-particle term is absent, i.e. if $\gamma=0$.
This is the case if
\be
\mbox{tr}\left[H\left(\sigma_z^{(1)}\otimes \sigma_z^{(2)}\otimes
\sigma_z^{(3)}\right)\right]=0.
\ee
This in turn is equivalent to
$\lambda_{000}+\lambda_{011}+\lambda_{110}+\lambda_{101}
=\lambda_{001}+\lambda_{010}+\lambda_{100}+\lambda_{111}.$
So any probability distribution $P\in\EEE_2$ fulfils 
$\ln[P(000)] +\ln[P(011)] + \ln[P(110)] + \ln[P(101)]=
\ln[P(001)]+\ln[P(010)]+\ln[ P(100)]+\ln[P(111)]$,
see also Ref.~\cite{amari01}. Equivalently, one has
\BE
&&P(000) \cdot P(011) \cdot P(110)\cdot P(101)\nonumber \\
&=&P(001) \cdot P(010) \cdot P(100) \cdot P(111).
\label{e2cond}
\EE
These considerations can now be used to show that the above two statements (1)
and (2) do not hold, and that instead local transformations do not leave
$\EEE_2$ invariant.  To this end one use Eq. (\ref{eq:alpha}) to generate
two-particle Hamiltonians and their associated probability distributions $P \in
\EEE_2$ at random and then subsequently apply a local transformation $T^{\rm
loc}$ with randomly chosen matrix elements $A^{(i)}_{\mu_i,\nu_i}$. Relation
(\ref{e2cond}) can then be used to check whether or not the outcome is still in
$\EEE_2$. It turns out that generally $T^{\rm loc}(P)\notin \EEE_2$. This proves
that $D_2$ can increase under local operations.

Our considerations complement those of Ref.~\cite{zhou09}, where it was observed
that a quantum analog of the quantity $D_k$ can increase under local operations
and classical communication. In the quantum case, however, the 
possible local transformations form a much larger class of maps: For instance,
the positive maps on a single
qubit are parameterized by 12 parameters, while in the classical case, a
transformation on a single particle
has only two degrees of freedom. Our example shows that the increase of $D_k$
occurs already in the classical regime and is also not due to possible
non-commuting terms in the Hamiltonian in the quantum setting. 

Finally, we would like to stress that the manifold $\EEE_1$ of distributions
factorizing over individual particles is clearly invariant under local
transformations. Hence, the quantity $D_1$ (also referred to as
multi-information \cite{amari01, multiinformation}) does not increase under
local transformations.

\subsection{Tracing out particles can increase $D_k$}
\begin{figure*}[t!!!]
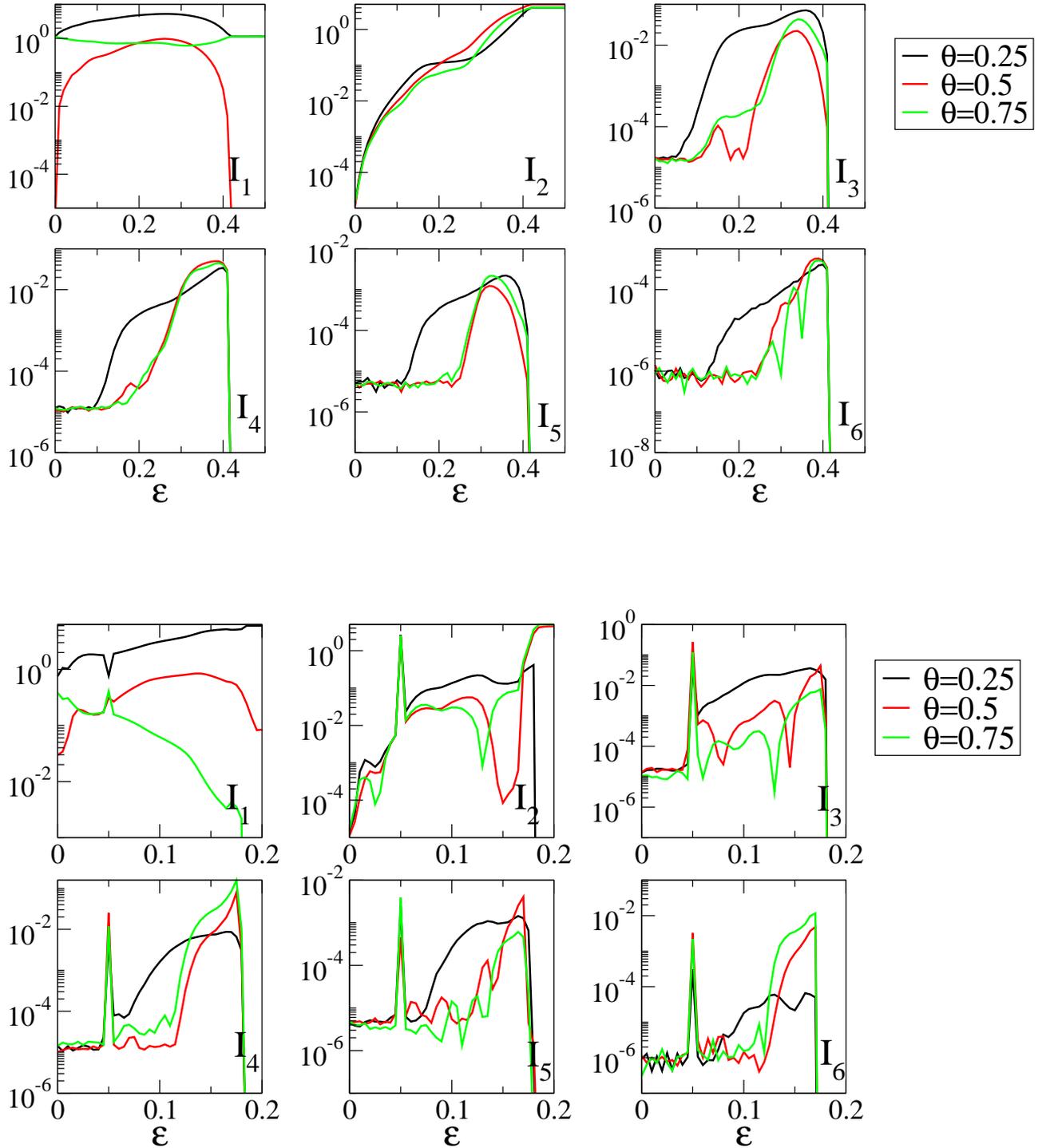

\centering
~~~~~~~~~~~\includegraphics[scale=0.65]{fig2a.eps}
\vspace{4em} \\
~~~~~~~~~~~\includegraphics[scale=0.725]{fig2b.eps}
\caption{(color online). Complexity measures $I_k$, $k=1,\dots,6$ as a function
of the coupling parameter $\varepsilon$ for a system of $N=6$ coupled tent maps
(upper six panels), and for $N=6$ coupled logistic maps (lower six panels). Each
panel shows three curves corresponding to coarse-graining procedures with
thresholds $\Theta=0.25, 0.5$ and $0.5$ respectively. All data is obtained from
running $1000$ iterations of the iterative projection algorithm. Curves
represent the average of $I_k$ over $20$ samples with independent random initial
conditions $x_i(t=0)\in [0,1]$.}
\label{fig:thresh1}
\end{figure*}
 
A second requirement one may ask of correlation measures is that they do not
increase when individual degrees of freedom are integrated out. Mathematically
formulated, the question is whether the $M$-particle marginals ($M<N$) of a an
$N$-particle distribution in $\EEE_k$ are still generated by a $k$-particle
Hamiltonian. For our later discussion it is important to realize that tracing
out individual particles can be understood as carrying out a local
transformation with specific parameters. More precisely, one may for example
choose to apply the local transformation defined by
\be
A^{(1)} = 
\begin{pmatrix}
1-a_1 & a_1 \\
a_1 & 1-a_1
\end{pmatrix}
\ee
and
\be
A^{(k)} = 
\begin{pmatrix}
1 & 0 \\
0 & 1
\end{pmatrix}
\ee
for $k= 2, ..., N$. For the first particle reports state $0$ to the central site
with probability $1-a_1$, and state $1$ with probability $a_1$ irrespective of
the actual state of the particle. The states of the remaining particles are
reported faithfully. If applied to an $N$-particle distribution $P(\cdot)$ the
transformation $T^{\rm loc} =\bigotimes_{i=1}^{N} A^{(i)}$ will result in a
distribution of the form
\be
T^{\rm loc}(P)(\bsigma)=R(\sigma_1)P'(\sigma_2,\dots,\sigma_N)
\ee
where $R(\sigma_1=0)=1-a_1$ and $R(\sigma_1=1)=a_1$, and where $P'(\cdot)$ is
the $N-1$-particle marginal of the original distribution $P(\cdot)$, i.e.,
$
P'(\sigma_2,\dots,\sigma_N)=\sum_{\sigma_1} P(\sigma_1,\sigma_2,\dots,\sigma_N).
$
Thus $T^{\rm loc}(P)\in \EEE_k$ if and only if $P'\in\EEE_k.$

Again, one can directly find counterexamples, which show that $D_k$ can increase
under tracing our particles. For instance, if one generates a random
four-particle distribution in $\EEE_2$, then, after tracing out one particle,
the marginal distribution is generally not found to be in $\EEE_2$.

\section{Incorporating local transformations in the correlation
measure}\label{sec:im}

In the last section we have seen that the measures $D_k$ can increase under
local transformations, and that they are hence lacking a desirable property of
correlation measures. To overcome this, we may replace the manifolds $\EEE_k$ by
their local orbits, 
\be \LL_k = \{T(P)\vert T \in \TT^{\rm loc}, P \in \EEE_k \}.  \ee 
The notation $\TT^{\rm loc}$ here indicates local transformations of the type
defined in Eqs.~(\ref{eq:tprod1},\ref{eq:tprod2}). The set $\LL_k$ is manifestly
invariant under local transformations, hence the quantity 
\be \label{eq:ck}
C_k(P) = \inf_{Q \in \LL_k} D(P \Vert Q) 
\ee 
is a correlation measure which is nonincreasing under local transformations. It
is also nonincreasing under tracing out particles, as this is a special case of
a local transformation. Given the invariance of $\EEE_1$ under local
transformations, the quantity $C_1$ coincides with the multi-information.

In the following we will first discuss some basic properties of this measure,
before we will describe how to approximate  $C_k$ numerically. In the next
session we will then apply this measure of complexity to the example of a set of
coupled chaotic maps, we will also provide a comparison with existing results on
the measure $D_k$ \cite{kahle09}.
\subsection{Properties of the set $\LL_k$ and the measure $C_k$}

Let us first discuss some properties of $\LL_k$. 
First we observe that for $N>4$ particles $\LL_2$ is a set of measure zero in
the space of 
all probability distributions. To see this one first notes that $\EEE_k$ is
characterized 
by $\sum_{i=1}^{k}{{N}\choose{i}}$ free parameters, this is the number of
possible terms 
in a Hamiltonian up to and including order $k$ (see
Section~\ref{localincrease}).  Furthermore,
local transformations are parameterized by two variables per particle, i.e. by
$2N$ parameters in total.  Given that $N$-particle distributions carry $2^N-1$
free parameters and that $2^{N}-1 >\sum_{i=1}^{2}{{N}\choose{i}} +2N$ for $N
\geq 5$  we conclude that the set $\LL_2$ 
is of measure zero in the space of all probability distributions. Similarly, one
can argue
that the $\LL_k$ for small $k$ are of measure zero for many particles.

For lower particle numbers the set $\LL_2$ does not cover the whole space of
distributions 
either. To see this, consider three particles and the probability distribution
$P(\cdot)$ with $P(000)=P(011)=P(101)=P(110)=1/4$, and $P(\bsigma)=0$ otherwise.
We know already that 
$D_2(P)=1$ is relatively large, hence it is a natural candidate to be outside of
$\LL_2.$ In order to show that this is indeed  the case, assume there exists a
local transformation $T \in \TT^{\rm (loc)}$ and a $Q \in \EEE_2$ such that $P =
T(Q).$ We parameterize $T=A^{(1)}\otimes A^{(2)}\otimes A^{(3)}$ 
via $A^{(k)}={1-a_k\;\; b_k\choose a_k \; 1-b_k}$ and we have
\begin{eqnarray}
P(001)& \geq &Q(000)(1-a_1)(1-a_2)a_3,
\nonumber
\\
P(010) &\geq &Q(000)(1-a_1)a_2(1-a_3),
\nonumber
\\
P(100)& \geq& Q(000)a_1(1-a_2)(1-a_3),
\nonumber
\\
P(111) &\geq& Q(000)a_1a_2a_3.
\end{eqnarray}
The distribution $P(\cdot)$ is constructed such that the left-hand-sides of
these equations vanish. Assume now that $Q(000)> 0$. In order to satisfy the
above inequalities we must have $a_k \in \{0,1\}$ for all $k.$ In a similar
manner, making the assumption $Q(111)>0$ leads to the requirement
$b_k\in\{0,1\}$ for all $k$.  

For cases in which $Q(000)$ and $Q(111)$ are both non-zero this solves the
problem, this includes in particular all distributions $Q(\cdot)$ with full
support. The local transformations are then such that any party applies one of
the following transformations: (i) they do not make any modification to the
state they observe ($a_k = b_k=0$), (ii) they always flip the binary symbol they
receive ($a_k = b_k=1$) or (iii) they always report the same result to the
central site no matter what input they receive (e.g., if $a_k=0, b_k=1$ they
always report $0$ and if $a_k=1, b_k=0$ they always report $1$). With these
transformations, however, one cannot obtain the probability distribution
$P(\cdot)$ from a distribution in $\EEE_2$
[in the case (iii) this is impossible as $P$ assigns positive probability to bit
strings in which the variable $\sigma_k$ takes either of the two values $0$ and
$1$].

It remains to address cases for which there are $\bsigma\in\Omega$ such that
$Q(\bsigma)=0$ (i.e. $Q$ is in the closure $\overline{\EEE_2}$, but not in
$\EEE_2$ itself). First, using the reasoning from above, 
one can directly see that if $Q(0,\sigma_2,\sigma_3)>0$ for some
$\sigma_2,\sigma_3$, 
then $a_1 \in \{0,1\}$ and if
$Q(1,\sigma_2,\sigma_3)>0$ for some $\sigma_2,\sigma_3$, then $b_1 \in \{0,1\}.$
This solves the problem for many further cases, 
for instance if there are two events $\sigma_1, \sigma_2, \sigma_3$ and
$\sigma_1', \sigma_2', \sigma_3'$ with $\sigma_k \neq \sigma_k'$ and
$Q(\sigma_1,\sigma_2,\sigma_3)>0$ as well as $Q(\sigma_1', \sigma_2',
\sigma_3')>0$.

The only remaining cases, where we cannot set any constraints on $a_1$ (or
$b_1$) are of the type
where  $Q(0,\sigma_2,\sigma_3)=0$ for all $\sigma_2,\sigma_3$ [or 
$Q(1,\sigma_2,\sigma_3)=0$ for
all $\sigma_2,\sigma_3$]. But for such cases the value of $a_1$ (or $b_1$) has
no effect on $T(Q),$ so, without changing the outcome of the transformation,
such a parameter can be set to either $0$ or $1$, again leading us to the
desired result. Finally, note that due to continuity reasons and the the fact
that $\overline{\EEE_2}$ is compact, $P$ has a finite distance to $\LL_2$. Using
the algorithm outlined in the next section we find a numerical distance of
$C_2(P)\approx 0.689$.

 \begin{figure*}[t!!!]
\vspace{0em}
\centering
\includegraphics[scale=0.5]{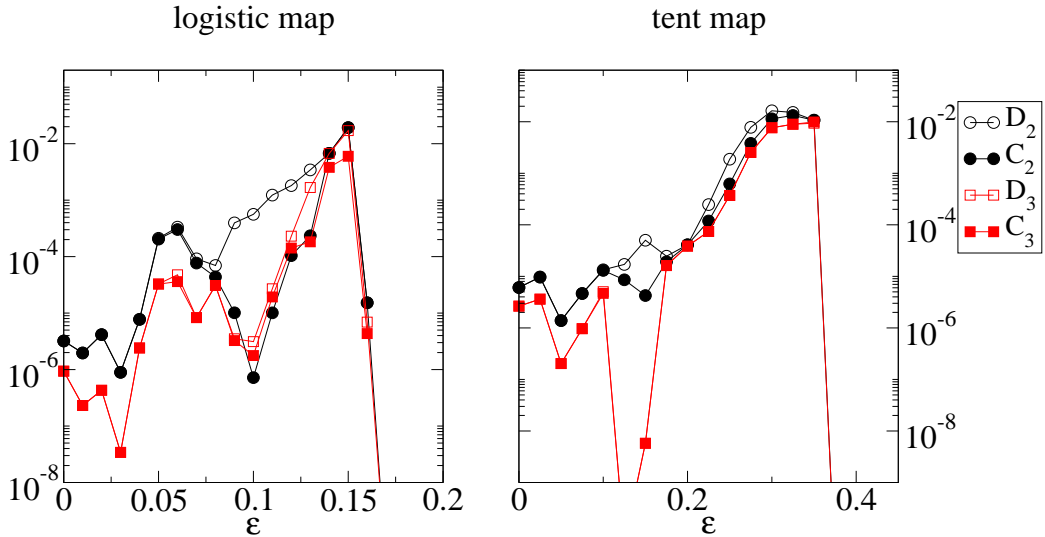}
\caption{(color online). Complexity measures $C_k$ and $D_k$ ($k=2,3$) for sets
of $N=4$ coupled logistic maps (left-hand panel), and $N=4$ coupled tent maps
(right-hand panel). Coarse-graining is performed at threshold $\Theta=1/2$. Open
symbols show the distances $D_k$ from exponential families $\EEE_k$ of
probability measures generated by $k$-particle Hamiltonians, filled symbols are
upper bounds on the distance from the local orbits of $\EEE_k$ (see text for
details).}
\label{fig:mod}
\end{figure*}

\subsection{Numerical calculation of $C_k$}
It is not a-priori not straightforward to compute $C_k(P)$ for a given
$N$-particle distribution $P(\cdot)$. We have however made some progress in
devising an iterative numerical scheme. We cannot prove at the moment that it
always converges, so strictly speaking it provides an upper bound on $C_k(P)$.
The quantity $C_k(P)$ as defined in Eq. (\ref{eq:ck}) is given by
\be \label{eq:ck2}
C_k(P) = \inf_{\begin{array}{c}T\in\TT^{(loc)}\\Q \in \EEE_k\end{array}} D[P
\Vert T(Q)],
\ee 
so we need to optimize the choice of $Q\in\EEE_k$ and that of $T\in\TT^{(loc)}$
simultaneously. We here proceed iteratively. Given a test distribution
$P(\cdot)$ we first find a Hamiltonian $H\in\QQ_k$, parameterized by a set
$\blambda^{(1)}$, so that $Q_{\blambda^{(1)}}$ is the best approximation of $P$
in $\EEE_k$. As a next step we then find parameters $\bomega^{(1)}$ which
minimize the distance $D[P\Vert T_{\bomega^{(1)}}(Q_{\blambda^{(1)}})]$, i.e. we
find the point in the local orbit of $Q_{\blambda^{(1)}}$ closest to $P$. We
then turn to an optimization of the $k$-particle Hamiltonian again, keeping
$T_{\bomega^{(1)}}$ fixed while optimizing $Q\in\EEE_k$. This is to say, find a
set of parameters $\blambda^{(2)}$ such that $D[P\Vert
T_{\bomega^{(1)}}(Q_{\blambda^{(2)}})]$ is minimized (subject to the constraint
$Q_{\blambda^{(2)}}\in\EEE_k$). Subsequently, we optimize the local
transformation again, and find a new set of parameters $\bomega^{(2)}$
minimizing $D[P\Vert T_{\bomega^{(2)}}(Q_{\blambda^{(2)}})]$. This procedure is
then iterated. Further details on the exact implementation can be found in the
Appendix.

\begin{figure*}[t]
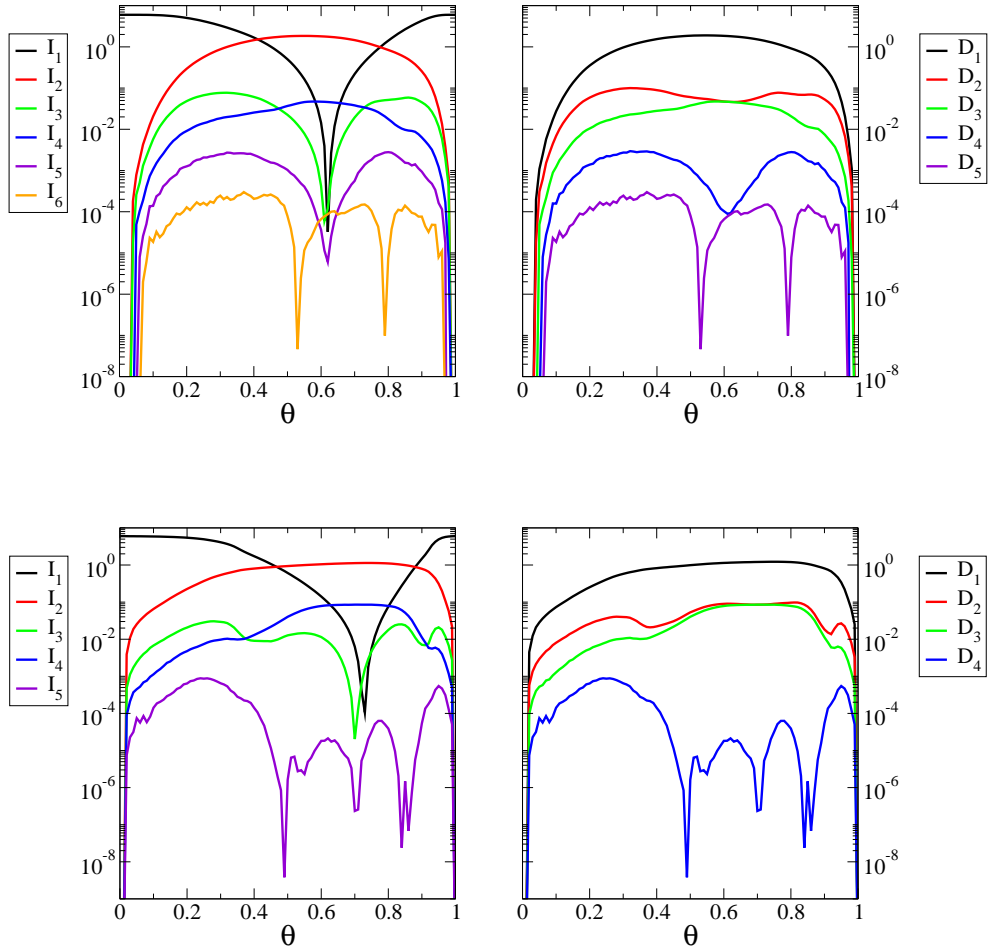

\vspace{0em}
\centering
\includegraphics[scale=0.5]{fig4a.eps}\\
\vspace{4em}
\includegraphics[scale=0.5]{fig4b.eps}
\caption{(color online). Complexity measures $I_k$ and distances $D_k$
($k=1,\dots,6$) in a system of $N=6$ coupled tent maps (upper two panels) and
for $N=6$ logistic maps (lower two panels) as functions of the coarse-graining
thresholds $\Theta$. The coupling strength is fixed at $\varepsilon=0.34$ for
the tent map and at $\varepsilon=0.17$ for the logistic map. All data is
obtained single runs of the coupled maps, started at a random initial condition.
The iterative projection algorithm is run for $1000$ iterations. }
\label{fig:thresh2}
\end{figure*}

\section{Application to coupled chaotic maps}\label{sec:log}
In this section we apply the proposed measures of complexity to a specific case
of an interacting multi-particle system. In particular we consider a coupled set
of $N$ discrete-time maps of the form \cite{kaneko}
\BE
x_i(t+1)&=&\left(1-\varepsilon\right)f[x_i(t)]\nonumber\\
&&+\frac{\varepsilon}{N-1}\sum_{j\neq i} f[x_j(t)],\label{eq:coupl}
\EE
where $i,j=1,\dots,N$. Similar arrangements were studied in Ref.~\cite{kahle09},
in particular different types of underlying adjacency networks were addressed.
We here focus on the all-to-all coupling described in Eq.~(\ref{eq:coupl}). The
parameter $\varepsilon\in[0,1]$ describes the strength of the interaction
between the maps. For $\varepsilon=0$ the $N$ maps are fully uncoupled, for
$\varepsilon>0$ they become increasingly more coupled. The chaotic maps we have
used are the tent map, given by 
\be
f(x)=\left\{\begin{array}{lcl} 2x & ~~~ & 0\leq x\leq 1/2 \\ 2-2x & ~~~ &
1/2\leq x\leq 1\end{array}\right.,
\ee
and the logistic map defined by
\be
f(x)=4x(1-x).
\ee
For both maps a symbolic dynamics of binary symbols is derived from the
continuous variables $x_i(t)\in[0,1]$ via a simple coarse graining procedure of
the type \cite{jalan}
\be
\sigma_i(t)=\left\{\begin{array}{lcl} 0 & ~~~ & x_i(t)\leq\Theta \\ 1 & ~~~ &
x_i(t)> \Theta\end{array}\right..
\ee
The parameter $\Theta\in(0,1)$ is here a coarse-graining threshold.

To obtain the probability distribution, we start with a random configuration of
the $x_i$ and interate 
the coupled map $10^6$ times. We disregard the first $10^5$ iterations and use
the remaining iterations
to determine our probability distribution.

\subsection{Previous results}
For completeness we will first re-iterate some of the results of
Ref.~\cite{kahle09}. Varying the value of the coupling strength $\varepsilon$ we
run simulations of a set of $N=6$ coupled tent and logistic maps. For the choice
$\Theta=1/2$ we obtain the corresponding symbolic dynamics, and measure the
resulting stationary distribution $P(\bsigma)$, where
$\bsigma=(\sigma_1,\dots,\sigma_6)$. We then use the iterative projection
algorithm proposed in Ref.~\cite{kahle09} (and discussed in more detail in the
Appendix), to obtain $D_k(P)$ for $k=0,\dots,6$. We then construct
$I_k=D_{k-1}-D_k$ ($k=1,\dots,6$) from the outcome of the iterative projection.
Results are shown in Fig.~\ref{fig:thresh1} (red curves). As reported in
Ref.~\cite{kahle09} the system of coupled maps is driven toward a state of
synchronized chaos if the coupling strength is sufficiently large ($\varepsilon
\gtrsim 0.45$ for the system of tent maps, $\varepsilon\gtrsim 0.2$ for the
logistic maps). In this regime the resulting distribution $P(\bsigma)$ is found
to be in $\EEE_2$, and accordingly we have $I^{(k)}=0$ for $k\geq 3$. With the
exception of $I^{(2)}$, which is seen to be monotonically increasing in
$\varepsilon$ all other measures $I^{(k)}$ attain maxima at $\varepsilon\approx
0.35$ for the system of coupled tent maps, indicating that the most complex
regime occurs just below the synchronization threshold \cite{kahle09}. A similar
observation is made for the system of logistic maps, see the red curves in the
lower six panels of Fig.~\ref{fig:thresh1}.

\subsection{Application of the modified complexity measure to coupled maps}
Results for the modified complexity measures $C_k$ are shown in
Fig.~\ref{fig:mod}.  Realistic computing resources at present only allow us to
study relatively small systems with $N=4$ particles, as our implementation of
the numerical computation of the improved measure $C_k$ is significantly more
demanding in computing time as that of the measures $D_k$. As the original
complexity measure  the modified measures are well able to detect the onset of
synchronization. This is unsurprising as $D_k=0$ implies $C_k=0$. 

For the cases we have studied here the functional dependence of $C_k$ broadly
follows that of $D_k$. 
For $k=3$ in particular our numerical results for the two measures are
essentially indistinguishable up to minor deviations. As seen in Fig.
\ref{fig:mod} the modified measure of complexity can on occasion deviate
substantially from the original one, see for example the data points near
$\varepsilon=0.1$ in the case of coupled logistic maps (left-hand panel of the
figure). It is here important to keep in mind that the algorithm for the
calculation  of $C_k$ is not guaranteed to converge. This, along with the
relatively high costs in computing time, are clear drawbacks of the complexity
measure we propose here. Future work may therefore address improved algorithms
for computing these measures, and/or more detailed comparison with the
complexity measures proposed by \cite{kahle09}. Should it turn out that both
sets of measures give broadly the same results, then it may well be appropriate
to continue to work with the $D_k$ (or $I_k$), despite the lack of a
non-increasing nature under local transformations.

\subsection{Dependence on single threshold}
As final point of our study we have investigated the influence of the
quantitative value of the coarse graining threshold $\Theta$ on the complexity
measures $D_k$ and $I_k$. It is here important to be aware that he coarse
graining procedure, mapping a real-valued variable $x_i(t)$ in the unit interval
onto a discrete degree of freedom $\sigma_i(t)\in\{0,1\}$, can be considered as
a local operation as well, but it 
is a local procedure in the construction of the probability distribution, and
not a local transformation applied after constructing the probability
distribution.\footnote{Therefore, one cannot set the strict requirement for any
complexity 
measure to be invariant under this transformation.}
In performing the coarse graining each particle is treated independently from
the rest of the system, and hence the applied operation is local.\footnote{The
following procedure would constitute an example of a nonlocal coarse-graining:
set $\sigma_i=0$ if $x_i(t)x_{i+1}(t)\leq1/2$ and $\sigma_i=1$ otherwise. The
expression $i+1$ is to be taken as `mod $n$'.}

As seen in Fig.~\ref{fig:thresh1} the behaviour of the $\{I_k\}$ as functions of
the coupling strength can depend considerably on the choice of the local
threshold. All curves, except for those of $I_2$ for the case of coupled tent
maps, show strong qualitative changes as $\Theta$ is varied. This is also
confirmed in
Fig.~\ref{fig:thresh2} where we show the $\Theta$-dependence of $I_k$ and $D_k$
explicitly at a fixed coupling strength $\varepsilon$, chosen below, but close
to the onset of synchronisation. Changing the numerical value of the threshold
can result in changes of the $I_k$ and $D_k$ by orders of magnitude, and as seen
in the left-hand panels of Fig.~\ref{fig:thresh2} it can also change the
relative ordering of the $I_k$, therefore affecting their interpretation as
measures of complexity. 

\begin{figure}[t!!!]
\vspace{0em}
\centering
\includegraphics[scale=0.45]{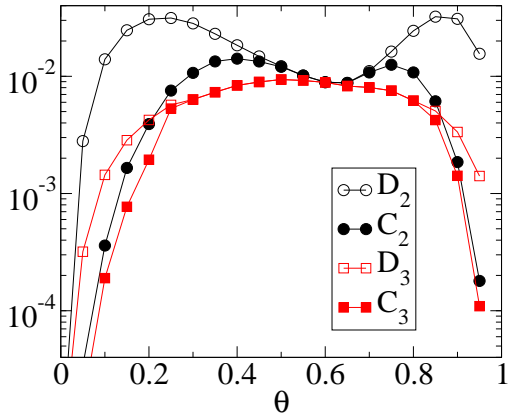}
\caption{(color online). Complexity measures $C_k$ and $D_k$ ($k=2,3$) for sets
of $N=4$ coupled tent maps at fixed coupling parameter $\varepsilon=0.34$ and at
varying coarse-graining threshold $\Theta$. Open symbols show the distances
$D_k$ from exponential families $\EEE_k$ of probability measures generated by
$k$-particle Hamiltonians, filled symbols are upper bounds on the distance from
the local orbits of $\EEE_k$.}
\label{fig:cthreshscan}
\end{figure}

In Fig.~\ref{fig:cthreshscan} we have tested the extent to which the use of the
modified complexity measures $C_k$ can remedy these effects. Again, for the
three-particle correlation, we find that using local orbits instead of
exponential families alone does not have any significant effect (we find that
$C_3\approx D_3$ for a broad range of values of the coarse-graining threshold
$\Theta$). The functional dependence of $C_2$ on the threshold parameter
$\Theta$ does however appear to be weaker than that of the measure $D_2$, we
detect more variation of $D_2$ for intermediate values of $\Theta$, where as the
numerical estimate of $C_2$ varies only by comparably small amount. While this
may provide some indication that the modified measures exhibit a lesser
dependence on the details of the coarse-graining procedure that the original
measure of complexity, more work is required to test whether this continues to
hold for other systems and/or systems with a larger number of particles

\section{Conclusion}\label{sec:con}

In summary, we have investigated measures of complexity and emergence for classical
interacting particle systems. Based on concepts from information geometry such
measures have recently been proposed \cite{kahle09}. The main idea behind these
measures is to consider the distance of a given probability distribution from
the set of distributions generated by $k$-particle Hamiltonians, that is to say
the set of distributions which can be factorized into a product of $k$-partite
distributions. We have shown that such measures are generally not invariant
under local transformations, acting on individual particles. In particular we
have demonstrated that, somewhat counterintuitively, local transformations can
increase the degree of multi-particle complexity in this measure, generating for
example $3$-particle correlations from $2$-particle correlations. Similarly
integrating out an individual degree of freedom can increase the complexity of
the remaining marginal. Recent work by Zhou \cite{zhou09} has revealed analogous
findings in the quantum realm, our work here demonstrates that these effects are
not intrinsically quantum, for example related to the non-commutative nature of
quantum mechanics. Instead they are also seen in classical systems. 

In order to remedy these undesired properties of existing complexity measures,
we have devised a modification, and propose to consider the distance from local
orbits of exponential families instead of the distance from the families
themselves. These orbits are manifestly invariant under local transformations of
the type we have defined, and as a consequence the resulting complexity measure
can only be reduced by applying a local transformation, but not increased. We
have devised a numerical scheme with which to calculate upper bounds for this
complexity measure, results are presented for dynamical systems composed of
multiple interacting chaotic maps. 

Finally we have investigated how the choice of local coarse-graining thresholds
affects the complexity measures proposed by Ref.~\cite{kahle09}. The coarse
graining procedure here constitutes a local manipulation as well, albeit
different from the type we consider above. Numerical results indicate that the
choice of coarse graining threshold can have significant effects on the
resulting estimates of complexity, hence care needs to be taken in turning the
underlying continuous degrees of freedom into symbolic language on the
coarse-grained level.

While the modified complexity measure we put forward here appears to exhibit
more convenient behaviour under local transformation that the ones proposed in
Ref.~\cite{kahle09} it is important for keep in mind that the algorithm we are
able to propose produces upper bounds for the required complexity measures, but
at present we are unable to decide whether or not it converges to the desired
quantities asymptotically. More work along these lines is required. Additionally
the numerical scheme we use here is limited to small systems, hence there is
significant room for improvement. This is however beyond the scope of the
current work which aims mainly to identify the key properties of the complexity
measures we have discussed, to relate them to measures based on exponential
families and to discuss their behaviour under local transformations. Our work is therefore mostly of a conceptual nature, and we hope our results will stimulate further work towards efficient algorithms for complexity measures with the local invariance propery. Additional
future lines of research may also address geometric concepts of distances other
than those based on the Kullback-Leibler divergence. Finally, it would also be
interesting to study the quantum analog of exponential families and their local
orbits in more detail. Work along these lines is in progress.

\begin{acknowledgments} 
We thank Nihat Ay, J\"urgen Jost, Robert MacKay, S\"onke Niekamp, and Eckehard
Olbrich for discussions. This work is supported by the Royal Society (reference
JP090467). TG would like to thank Research Councils UK for support (RCUK
reference EP/E500048/1). OG acknowledges support by the FWF (START prize
Y376-N16) and the EU (Marie Curie CIG 293993/ENFOQI).
\end{acknowledgments}
\vspace{1em}

\section{Appendix: Algorithms}
This Appendix contains details on the two main algorithms used in this work. In
Sec. \ref{sec:cipi} we provide the precise steps with which to compute $D_k(P)$
for a given distribution $P$, i.e. the distance from the set of distributions
generated by $k$-particle Hamiltonians. In Sec. \ref{sec:algoloc} we explain in
more detail how to compute an upper bound for $C_k(P)$, the distance of $P$ from
the local orbit of $\EEE_k$. 
\subsection{Iterative projection algorithm}\label{sec:cipi}
Our presentation here closely follows that of Ref.~\cite{cipi}, see also
Ref.~\cite{kahle09}. Given an (empirical) probability distribution $P(\cdot)$ on
$\Omega=\{0,1\}^N$, and an integer $k\leq N$ (the order of interaction at which
we are approximating), the algorithm to find $D_k(P)$ is as follows:
\begin{enumerate}
\item[1.] {\em Computation of marginals}: For each subset $A\subset V$ with $k$
elements ($|A|=k$), compute the following
\be
\alpha_A(\bsigma)=\sum_{\bsigma':\pi_A(\bsigma')=\pi_A(\bsigma)}P(\bsigma').
\ee
Here $\pi_A(\bsigma)$ denotes the projection of $\bsigma$ onto $A$, i.e.
$\pi_A(\bsigma)=(\sigma_i)_{\{i\in A\}}$. The quantity $\alpha_A(\bsigma)$ only
depends on the components $\sigma_i$ of $\bsigma$ with index $i\in A$, i.e. on
the $k$-variables $(\sigma_i)_{i\in A}$.
\item[2.] {\em Initialisation:} Initialise $Q$ as the flat distribution over
$\Omega$: $Q(\bsigma)=1/(2^N)$ for all $\bsigma$.
\item[3.] {\em Improve current approximation:} Run through all sets $A\subset V$
with $|A|=k$ and update $Q(\cdot)$ as follows:
\be
Q^{\mbox{\small new}}(\bsigma)=c_{A}(\bsigma)Q(\bsigma),
\ee
where
\be
c_A(\bsigma)=\frac{\alpha_A(\bsigma)}{\sum_{
\bsigma':\pi_A(\bsigma')=\pi_A(\bsigma)}Q(\bsigma')}.
\ee
\item[4.] {\em Update and iterate:} Replace $Q(\cdot)\rightarrow Q^{\mbox{\small
new}}(\cdot)$ and goto 3.
\end{enumerate}

\subsection{Approximate scheme to calculate distance from a local
orbit}\label{sec:algoloc}
 
Given a test distribution $P(\cdot)$ the algorithm proceeds as follows:
\begin{enumerate}
\item[1.] Find a Hamiltonian $H\in\QQ_k$, parameterized by a set
$\blambda^{(1)}$, so that $Q_{\blambda^{(1)}}=Z^{-1}\exp(H)$ is the best
approximation of $P$ in $\EEE_k$. This can for example be done using the
iterative projection algorithm described above.
 \item[2.] {\em Optimization of the local transformation for a given
Hamiltonian:} Given a Hamiltonian parameterized by $\blambda^{(m)}$ find
parameters $\bomega^{(m)}$ which minimize the distance $D[P\Vert
T_{\bomega^{(m)}}(Q_{\blambda^{(m)}})]$, i.e. we find the point in the local
orbit of $Q_{\blambda^{(m)}}$ closest to $P$. The optimization of the parameter
set $\bomega=(a_1,b_1,\dots,a_N,b_N)$ is carried out in an iterative manner,
i.e. we first optimize $a_1$, then $b_1$, then $a_2$, and so on, keeping
previously optimized parameters fixed. This procedure is iterated a number of
times (typically in excess of $10$ sweeps). Each parameter optimization is
carried out using an iterative Monte Carlo procedure, based on first randomly
choosing test values in the interval $[0,1]$ and then choosing a sequence of
nested intervals with decreasing the range from which more finely spaced
subsequent test values are drawn.
\item[3.] {\em Optimization of the Hamiltonian given a local transformation:}
Given a local transformation defined by parameters $\bomega^{(m)}$  find a set
of parameters $\blambda^{(m+1)}$ such that $D[P\Vert
T_{\bomega^{(m)}}(Q_{\blambda^{(m+1)}})]$ is minimized (subject to the
constraint $Q_{\blambda^{(m+1)}}\in\EEE_k$). Similar to the procedure outlined
above for the optimization of $\bomega$ we optimize one of the parameters
$\blambda$ at a time, carrying out a set of typically $10-20$ or more sweeps
over all parameters. Any one parameter is optimized by applying an deterministic
search algorithm on the interval $[-10,10]$ (with a suitable discretization).
The range of each parameter is therefore effectively truncated. The constraint
$Q_{\blambda^{(m+1)}}\in\EEE_k$ is taken into account by pre-setting the
coefficients of $\ell$-particle terms, $\ell>k$ to zero. For a three-particle
system for example we have
$\blambda=(\alpha_1,\alpha_2,\alpha_3,\beta_{12},\beta_{23},\beta_{13},\gamma)$,
see Eq. (\ref{eq:alpha}). If the wish to compute differences say from $\EEE_2$
we would set $\gamma=0$ from the start, and only optimize the remaining entries
in $\blambda$.
\item [4.] This procedure is then iterated, i.e., goto 2.
\end{enumerate}
We would like to stress that no claim is made that this algorithm provides an
exact result for $C_k$, this may not even be the case in the limit of an
infinite number of iterations. The numerical scheme is an approximate procedure,
providing at least an {\em upper bound} on $C_k$. It is also interesting to note
that the algorithm contains a certain stochastic element (rooted in the
Monte-Carlo optimization of the parameters $\bomega$ as described above). We
have found that it can be beneficial to allow occasional increases in the
estimates of distance (i.e. `uphill' motion), as this prevents dynamical arrest
in local minima. Results reported are the minimal estimate of distance obtained
during any one run of the algorithm, and not necessarily the distance estimate
at the end of the nested set of iterations.


\end{document}